\documentstyle[12pt]{article}

\def\eg{{\it e.g. }}

\title{ {\bf Dissipative N -- body code for galaxy evolution.}
        \bigskip      
      }

\author{ {\bf Peter Berczik \& Sergei G. Kravchuk} \bigskip \\ 
         Main Astronomical Observatory of \\
         Ukrainian National Academy of Sciences \\
         252650, Golosiiv, Kiev-022, Ukraine \bigskip \\
         e-mail: {\tt berczik@mao.kiev.ua}
         \bigskip
       }

\date{September 1, 1998}

\begin{document}

\maketitle              

%----------------------------------------------------------------------------%

\begin{abstract}

    The evolving galaxy is considered as a system of baryonic fragments 
embedded into the static dark nonbaryonic (DH) and baryonic (BH) halo 
and subjected to gravitational and viscous interactions. Although the 
chemical evolution of each separate fragment is treated in the frame of 
one -- zone close box model with instantaneous recycling, its star 
formation (SF) activity is a function of mean local gas density and,
therefore, is strongly influenced by other interacting fragments. In 
spite of its simplicity this model provides a realistic description of 
the process of galaxy formation and evolution over the Hubble timescale.

\end{abstract}

%----------------------------------------------------------------------------%

%\Large{
%\baselineskip=24pt

%----------------------------------------------------------------------------%

%\eject

\section{Introduction}

     Recent advances in extragalactic astrophysics show the close link 
of a disk galaxy dynamical evolution and its chemical and photometric 
behaviour over the Hubble timescale. In spite of remarkable succes of 
the modern theory of galaxy chemical evolution in explaining the 
properties of evolving galaxies (\cite{P94}) its serious shortcomings 
concern the multiparameter character and practical neglecting of 
dynamical effects. The inclusion of simplified dynamic into the chemical 
network (\cite{SH96}) and vice versa the inclusion of simplified 
chemical scheme into the sophisticated 3D hydrodynamical code 
(\cite{SM94,BK97}) gives very promising results and allows to avoid a 
formal approach typical to standard theory.

     In this paper the interplay between a disk galaxy dynamical 
evolution and its chemical behaviour is studied in a frame of a simplified 
model which provides a realistic description of the process of galaxy 
formation and evolution over the cosmological timescale.

%----------------------------------------------------------------------------%

%\eject

\section{Initial conditions}

     The evolving galaxy is treated as a system of baryonic fragments 
embedded into the extended halo composed of dark nonbaryonic and 
baryonic matter. The halo is modelled as a static structure with dark 
(DH) and diluted baryonic (BH) halo components having Plummer -- type 
density profiles (\cite{DC95}):

$$
\rho_{BH}(r) = \frac{ M_{BH} }{ \frac{4}{3} \pi b_{BH}^3 } \cdot 
               \frac{ b_{BH}^5 }{ (r^2 + b_{BH}^2 )^{\frac{5}{2}} }
$$

and

$$
\rho_{DMH}(r) = \frac{ M_{DMH} }{ \frac{4}{3} \pi b_{DMH}^3 } \cdot 
                \frac{ b_{DMH}^5 }{ (r^2 + b_{DMH}^2 )^{\frac{5}{2}} },
$$

where

$$
M_{DMH} = 10^{12} \; M_\odot, \; b_{DMH} = 25 \; {\rm kpc}, \;
$$
$$
M_{BH} = 10^{11} \; M_\odot, \; b_{BH} = 15 \; {\rm kpc}.
$$ 

     The dense baryonic matter (future galaxy disk and bulge) of total 
mass of $ M_{gas} = 10^{11}\; M_\odot $ is assumed to be distributed 
among $ N = 2109 $ particles -- fragments. The single particle density 
profile is also assumed to be a Plummer one. Its mass is taken to be $ 
m_i = M_{gas}/N $ and radius $ h_i = 1 $ kpc. Initially all particles 
are smoothly placed inside the sphere of radius $ R_{gas} = 50 $ kpc and 
are involved into the Hubble flow ($ H_{0} = 75 $ km/s/Mpc) and solid -- 
body rotation around $ z $ axis. The initial motion of this system is 
described as:

$$
{\bf V}(x, y, z) = [{\bf \Omega}(x, y, z) \times {\bf r}]
                   + H_{0}\cdot {\bf r} + {\bf DV}(x, y, z),
$$

where $ {\bf \Omega}(x,y, z) = (0, 0, 1) \cdot \Omega_{cir} $ is an 
angular velocity of the rotating sphere, $ \Omega_{cir} = 
V_{cir}/R_{gas} $ and

$$
V_{cir} = \sqrt{ G \cdot \frac{ M_{gas} + M_{DMH} + M_{BH} }{R_{gas} } }.
$$

The components $ DV_x $, $ DV_y $, $ DV_z $ of the random velocity $ 
{\bf DV} $ vector are assumed to be initially randomly distributed 
within an interval $ 0 \div 10 $ km/s.

%----------------------------------------------------------------------------%

%\eject

\section{N -- body code}

     The dynamical evolution of baryonic matter fragments which are 
subjected to gravitational influences of DM--baryonic halo and 
interfragment interactions is followed by means of effective  N -- body 
integrator with individual time step. The dynamics of such N -- body 
system is described by following equations:

\begin{equation}
    \left\{
    \begin{array}{lll}
    d{\bf R}_i/dt = {\bf V}_i,                   \\
                                                 \\
    d{\bf V}_i/dt = {\bf A}_i({\bf R}, {\bf V}). \\
    \end{array}
    \right.
\label{eq:def_dr_dv}
\end{equation}

     The acceleration of $ i $ - th particle $ {\bf A}_i $ is defined as 
a sum of three components.

\begin{equation}
{\bf A}_i = {\bf A}^{INT}_i + {\bf A}^{EXT}_i + {\bf A}^{VISC}_i,
\label{eq:A1}
\end{equation}

     where the first term $ {\bf A}^{INT}_i $ accounts for gravitational 
interactions between fragments. The second one $ {\bf A}^{EXT}_i $ is 
defined as an external gravitational acceleration caused by the DM and 
baryonic halo. The last term $ {\bf A}^{VISC}_i $ corresponds to the 
viscous decceleration of fragment when passing through the baryonic 
halo.

     The gravitational interaction between fragments is defined as the 
interaction of $ N $ Plummer profile elements:

\begin{equation}
{\bf A}^{INT}_i = - G \cdot \sum_{j=1, j \ne i}^{N} \frac{ m_j }
                 { ({\bf R}^2_{ij} + h^2_{ij})^{\frac{3}{2}} }
                 \cdot {\bf R}_{ij}.
\label{eq:A_int}
\end{equation}

     Here $ h_{ij} = (h_i + h_j)/2 $ and $ {\bf R}_{ij} = {\bf R}_i - 
{\bf R}_j $.

     Accounting for that the halo DM and baryonic components are also 
Plummer spheres the second term becomes:

\begin{equation}
{\bf A}^{EXT}_i = - G \cdot ( \frac{ M_{DMH} }
                    { ({\bf R}^2_{i} + b^2_{DMH})^{\frac{3}{2}} }
                  + \frac{ M_{BH} }
                    { ({\bf R}^2_{i} + b^2_{BH})^{\frac{3}{2}} } )
                    \cdot {\bf R}_{i}.
\label{eq:A_ext}
\end{equation}

     The form of the last term $ {\bf A}^{VISC}_i $ will be discussed in 
the next subsection.

      The characteristic time step $ \delta t_i $ in the integration 
procedure for each particle is defined as:

\begin{equation}
\delta t_i = Const \cdot \min_{j} [ 
       \sqrt{ \frac{ \mid {\bf R}_{ij} \mid }{ \mid {\bf A}_{ij} \mid } },
       \frac{ \mid {\bf R}_{ij} \mid }{ \mid {\bf V}_{ij} \mid }
                ]
\label{eq:def_dt}
\end{equation}

     where the $ {\bf V}_{ij} = {\bf V}_i - {\bf V}_j $ and the $ {\bf 
A}_{ij} = {\bf A}_i - {\bf A}_j $. 

     Here the $ Const $ is a numerical parameter equal to $ Const = 
10^{-2} $ that provides a nice momentum and energy conservation over the 
integration interval of about $ 15 $ Gyr. For example, in the 
conservative case (when viscosity of the system is set equal to $ 0 $), 
the final total error in the energy equation is less than $ 1 \% $.

%----------------------------------------------------------------------------%

%\eject

\section{Viscosity model}

     The viscosity term $ {\bf A}^{VISC}_i $ is artificially introduced 
into the model so as to match the results of more sophisticated SPH 
approach on dynamical evolution of disk galaxies. The best fitness of 
results of this simplified approach with SPH modelling data (see \eg 
\cite{BK97}) is achived when the momentum exchange between the 
baryonic halo and moving particles is modelled by the following 
expression:

\begin{equation}
{\bf A}^{VISC}_i = - {k\cdot\bf V}_i \cdot
                     \frac{\mid {\bf V}_i \mid}{R_{VISC}} \cdot
                     \frac{\rho_{BH}(r)}{\rho_{VISC}} \cdot
                     \frac{m^{gas}_i}{m^{gas}_i +m^{star}_i}.
\label{eq:A_visc}
\end{equation}

     Here $ \rho_{VISC} $ and $ R_{VISC} $ are numerical parameters set 
equal to $ \rho_{VISC} = 0.2 $ cm$^{-3}$ and $ R_{VISC} = 10 $ kpc. The 
vector $ {\bf V}_i $ is a particle velocity vector. It is to be noted 
that single particle is assumed to have a total mass which doesn't 
change with time and is defined as sum $ m_i \equiv m^{gas}_i + 
m^{star}_i $. But masses of its gas and star components $ m^{gas}_i $ 
and $ m^{star}_i $ are variable values and are defined by the temporal 
evolutionary status of the given fragment. Initially $ m^{star}_i = 0 $. 
The results of fitness show that for $ z $ component of viscosity term $ 
k = 1 $. In the galaxy plane where it is necessary to account for 
baryonic halo and baryonic fragments partial corrotation the dynamical 
friction is decreased and for $ x, y $ component of viscosity term $k$ 
reduced to the value $ 0.15 $. 

%----------------------------------------------------------------------------%

%\eject

\section{Density definition}

     In the frame of the multifragmented model the definition of local 
gas density is introduced in the SPH manner, e.g. local gas density 
depends on the total mass of matter contained in the sphere of radius $ 
H_i $ around the $ i $ -- th particle. For each $ i $ -- th particle the 
value of its smoothing radius $ H_i $ is chosed (using the quicksort 
algorithm) requiring that the volume of such radius compraises $ N_B = 
21 $ nearest particles (i.e. $ \approx $ 1 \% of total number of 
particles $ N $). Therefore, the total mass $ M_i $ and density of gas $ 
\rho_i $ inside this sphere are defined as

\begin{equation}
M_i = \sum_{j=1}^{N} \Delta m^{gas}_{ij}, \; \; \; \; \; \rho_i = \frac{M_i}{ \frac{4}{3} \pi H_i^3 },
\label{eq:Mass_i}
\end{equation}

     where $ \Delta m^{gas}_{ij} $ is defined as
     
$$      
if \;\;\; \mid {\bf R}_{ij} \mid > (H_i + h_j) => \Delta m^{gas}_{ij} = 0,  
$$
$$
if \;\;\; \mid {\bf R}_{ij} \mid < (H_i - h_j) => \Delta m^{gas}_{ij} = m^{gas}_j, 
$$      
$$
else \;\;\; \Delta m^{gas}_{ij} = m^{gas}_j \cdot \frac{H_i + h_j - \mid {\bf R}_{ij} \mid}{2 \cdot h_j}. 
$$      

%----------------------------------------------------------------------------%

%\eject

\section{Star formation and SN explosions}

     A forming disk galaxy is modelled as a system of interacting 
fragments (named as particles) embedded into the extended halo. Each 
particle is composed of gas and stellar components and its total mass is 
defined as $ m_i \equiv m^{gas}_i + m^{star}_i $. Initially all 
particles are purelly gaseous and, therefore, initially $ m^{star}_i = 0 
$. To follow a particle star formation (SF) activity a special timemark 
$ t^{begSF}_i $ is introduced which initially is set equal to $ 
t^{begSF}_i = 0 $. The particles eligible to star formation events are 
chosen as particles which still have a sufficient amount of the gas 
component and their densities exceed some critical value $ \rho_{minSF} 
$ during some fixed time interval $ \Delta t_{SF} $ (of order of 
free--fall time): 

\begin{equation} 
\left\{ 
\begin{array}{lllll} 
\rho_i > \rho_{minSF},                          \\
\\ 
t_i - t^{begSF}_i > \Delta t_{SF},              \\ 
\\ 
m^{gas}_i/(m^{gas}_i + m^{star}_i) > 10^{-4}.   \\ 
\end{array} \right. 
\label{eq:SF_yes} 
\end{equation}

     Here $ \Delta t_{SF} = 50 $ Myr, and $ \rho_{minSF} = 0.01 $ 
cm$^{-3}$ (this last value is not crucial and is only limiting one).

      If the particle was subjected to SF activity the parameter $ 
t^{begSF}_i $ is set equal to $ t^{begSF}_i = t_i $, , and $ m_i^{star} 
$ and $ m_i^{gas} $ are redefined as

\begin{equation}
    \left\{
    \begin{array}{lll}
    m_i^{star} := \epsilon \cdot (1 - R) \cdot m_i^{gas} + m_i^{star}, \\
                                                                       \\
    m_i^{gas}  := (1 - \epsilon \cdot (1 - R) ) \cdot m_i^{gas}.       \\
    \end{array}
    \right.
\label{eq:m}
\end{equation}

     Here $ \epsilon $ is a SF efficiency which is defined as
     
\begin{equation}
\epsilon = \alpha \cdot \frac{\rho_i}{\rho_{SF}} \cdot (1 - exp(-\frac{\rho_{SF}}{\rho_i})).
\label{eq:epsilon}
\end{equation}

     Therefore, 
     
$$      
if \;\;\; \frac{\rho_i}{\rho_{SF}} \rightarrow 0 => \epsilon \rightarrow \alpha \cdot \frac{\rho_i}{\rho_{SF}},  
$$
$$
if \;\;\; \frac{\rho_{SF}}{\rho_i} \rightarrow 0 => \epsilon \rightarrow \alpha.
$$      

          To match available observational data on star formation 
efficiency (see \eg\cite{WL83}) parameters $ \alpha $ and $ \rho_{SF} $ 
are set equal to $ \alpha = 0.5 $ and $ \rho_{SF} = 10 cm^{-3} $.
The chemical evolution of each separate fragment is treated in the frame 
of one -- zone close box model with instantaneous recycling. In the 
frame of this approach for the returned fraction of gas from evolved 
stars is taken a standard value $ R = 0.25 $.

     Following the instantenuous recycling approximation it is assumed 
that after each SF and SN explosions the heavy element enriched gas is 
returned to the system and mixed with old (heavy element deficient) gas. 
After each act of SF and SN explosions the value of heavy 
element abundances of gas in particle is upgraded according to:

\begin{equation} 
Z_i := Z_i + 
\frac{\epsilon \cdot R \cdot \Delta Z} {1 - \epsilon \cdot (1-R)}. 
\label{eq:def_Z} 
\end{equation}

     The value $ \Delta Z = 0.01 $ is used as an average value for all 
values of $ Z = 0.001 \div 0.04 $ (see \cite{BK97}). Initially $ Z_i = 
0.0 $ in all particles.

%----------------------------------------------------------------------------%

%\eject

\section{Conclusion}

     The proposed simple model provides the self - consistent picture of 
the process of galaxy formation, its dynamical and chemical evolution is 
in agreement with the results of more sophisticated approaches (see \eg 
\cite{SM94, RVN96, SH96, BK97}).

\begin{itemize}

   \medskip
   
   \item The rapidly rotating protogalaxy finally formed a three - 
component system resembling a typical spiral galaxies: a thin disk and 
spheroidal component made of gas and/or stars and a dark matter halo. 

     Fig.1 and Fig.2 show respectively the star formation rate and the 
galaxy total stellar mass as a function of time. Fig.3 shows the 
cylindrical distribution of stellar (upper curve) and gaseous (lower 
one) components of the final model disk galaxy as a function of a 
distance from the galaxy center in the galactic plane. 

    The total star formation rate (SFR) is a succesion of short bursts 
which doesn't exceed $ 28 \; M_{\odot}$/yr. During first $ 2 $ Gyr of 
evolution only about $ 20 \% $ of total galaxy mass is transforms into 
the stars. The SFR gradually decreases, during the further evolution, to 
the value of about $ 2 \; M_{\odot} $/yr typical for our own Galaxy. The 
final total stellar and gas mass of the model galaxy disk are about $ 92 
\% $ and $ 8 \% $. All these data as well as surface densities 
distributions of stellar and gaseous components (see Fig.4) are in nice 
agreement with present date observational data (\cite{KG89, P94}).  

   \medskip
   
   \item The disk component posesses a typical spiral galaxy rotation 
curve and the distribution of radial and $ V_z $ - th velocities of 
baryonic particles clearly show the presence of the central bulge (see 
Fig.6 and Fig.7).
   
   \medskip
   
   \item The metallicities and the global metallicity gradient resemble 
distributions observed in our own Galaxy (Fig.5). The averaged observed 
value of global metallicity $ Z/Z_\odot(r) $ (see \cite{P94}) is shown
in this Fig. as a solid line.
   
\end{itemize}

%----------------------------------------------------------------------------%

%\eject

     {\bf Acknowledgements:} Peter Berczik would like to acknowledge the 
American Astronomical Society for financial support of this work under 
International Small Research Grant.
     
%----------------------------------------------------------------------------%

%\eject

\section{Figures}

\noindent Figure 1: The variation of total star formation rate  
of forming disk galaxy with time.

\bigskip

\noindent Figure 2: The growth of galaxy stellar mass with time. 

\bigskip

\noindent Figure 3: The cylindrycal distribution of masses of stellar 
(upper curve) and gaseous (lower one) components as a function of 
distance from galaxy center in a galactic plane.

\bigskip

\noindent Figure 4: The surface density radial distributions of 
stellar and gaseous components (stellar component is shown by filled 
stars, gaseous one by astericks). Theoretical distributions (not
scaled) of surface density for radial exponential scale lengthes 
$ 2.0 $ and $ 3.0 $ kpc are shown below by lines.  

\bigskip

\noindent Figure 5: The radial distribution of heavy element abundances 
(averaged observed distribution of $ z $ for our Galaxy is shown by 
solid line).

\bigskip

\noindent Figure 6: The final distribution of $ V_z $ - th velocities
of baryonic gas--stellar particles.

\bigskip

\noindent Figure 7: The final galaxy rotation curve.

%----------------------------------------------------------------------------%

%}   %%% end of Large

%\eject

%\small{

%}   %%% end of small

%----------------------------------------------------------------------------%

%############################################################################%
\eject

%\begin{figure*}[t]
\begin{figure*}[htbp]

\vspace{20.0cm}
\includegraphics{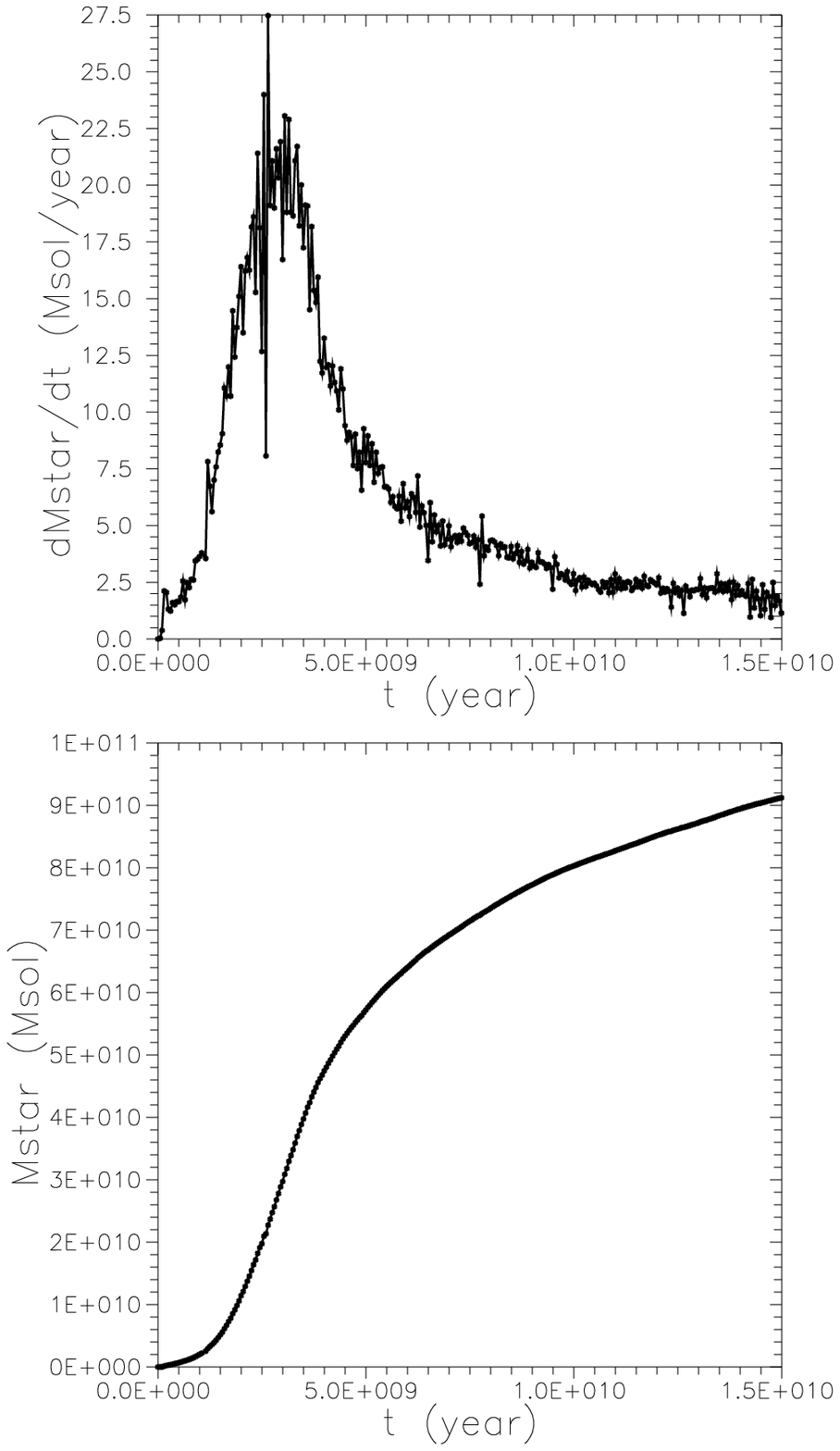}

%\caption{} 
 
%\label{fig1}
\end{figure*}
%############################################################################%

\eject

%\begin{figure*}[t]
\begin{figure*}[htbp]

\vspace{20.0cm}
\includegraphics{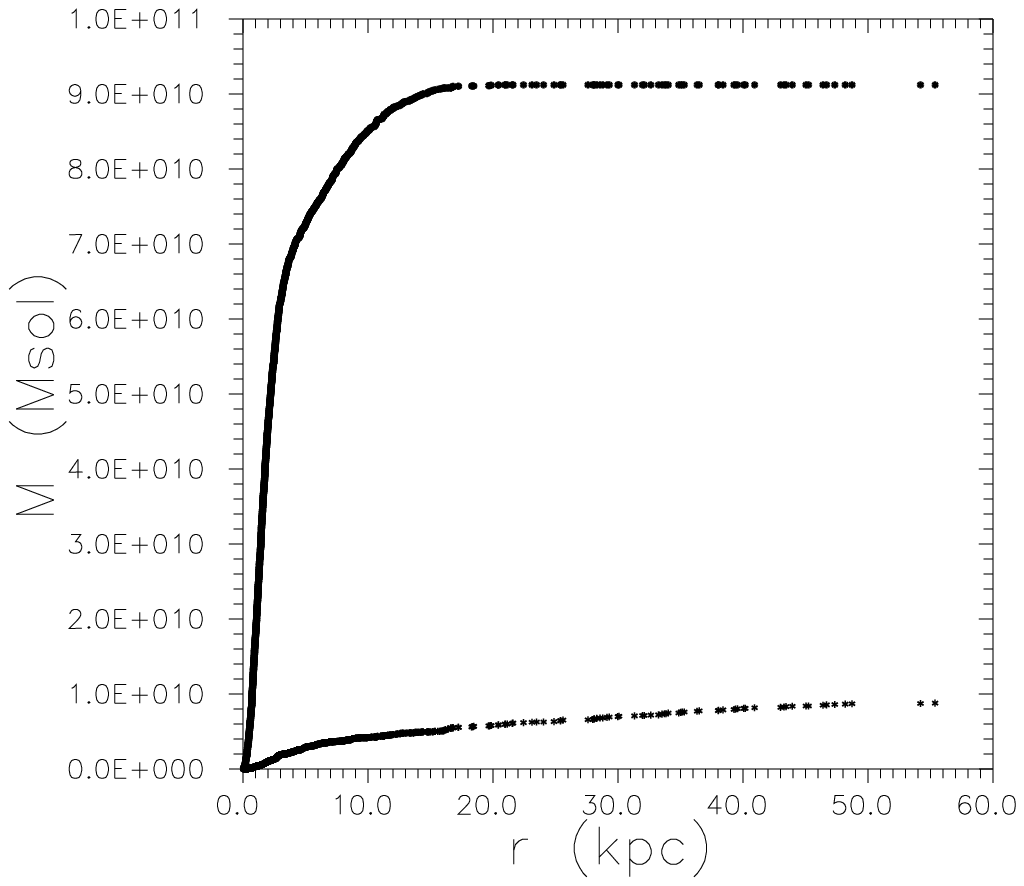}

%\caption{}
 
%\label{fig2}
\end{figure*}
%############################################################################%

\eject

%\begin{figure*}[t]
\begin{figure*}[htbp]

\vspace{20.0cm}
\includegraphics{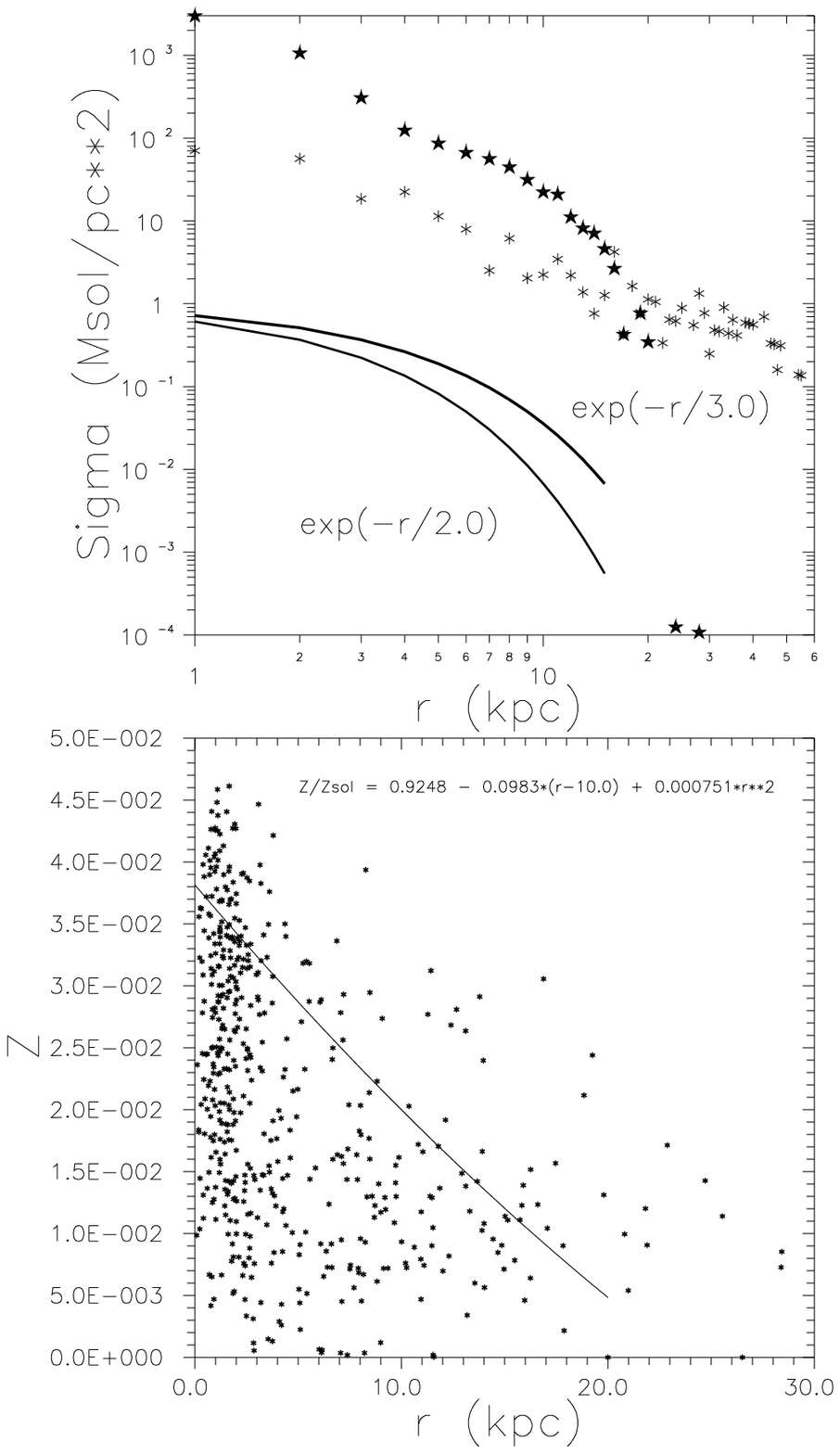}

%\caption{}
 
%\label{fig3}
\end{figure*}
%############################################################################%

\eject

%\begin{figure*}[t]
\begin{figure*}[htbp]

\vspace{20.0cm}
\includegraphics{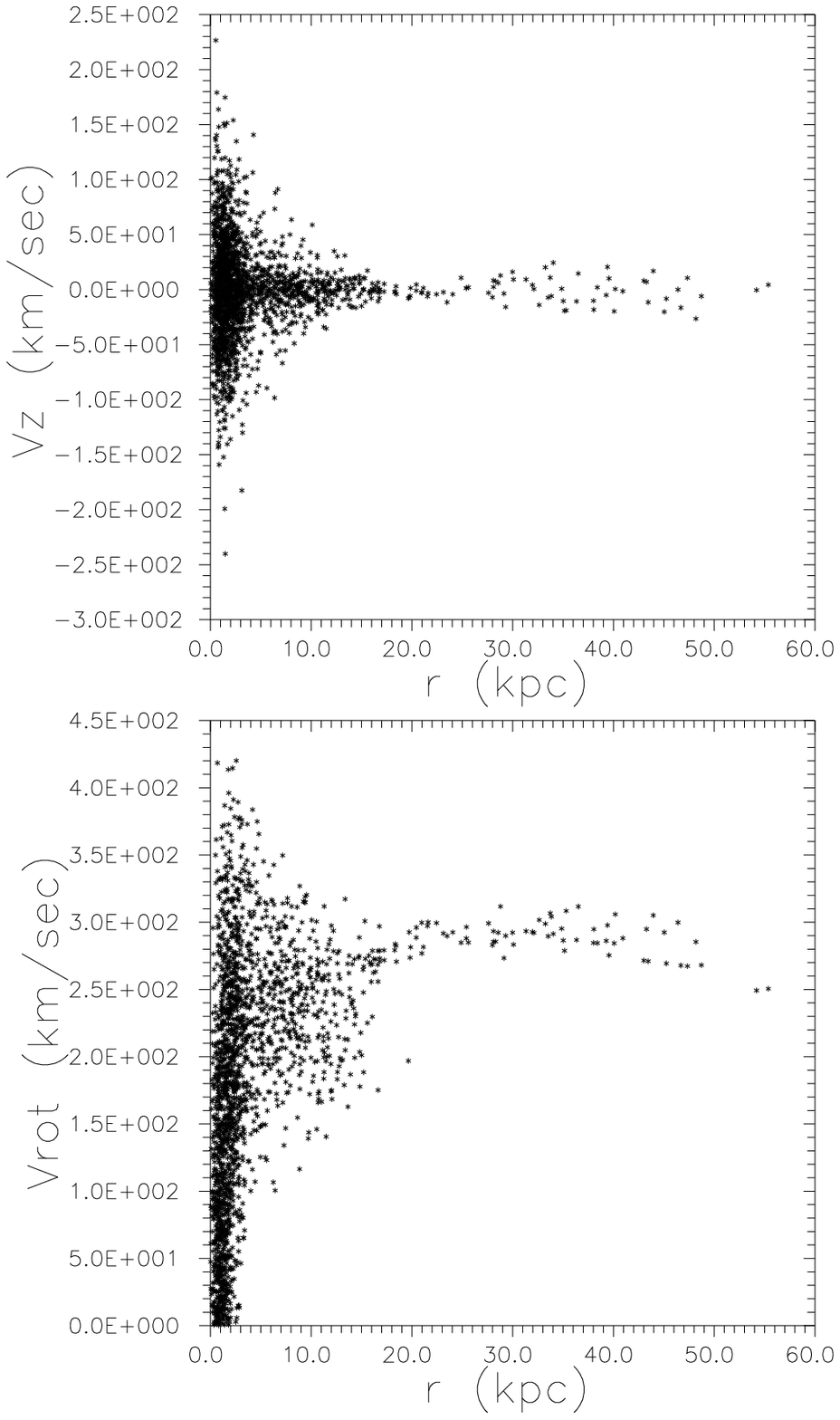}

%\caption{}
 
%\label{fig4}
\end{figure*}
%############################################################################%

%----------------------------------------------------------------------------%

\end{document}